\documentclass[english,reprint,aps,prb,superscriptaddress,longbibliography]{revtex4-2}
\usepackage[T1]{fontenc}
\usepackage[latin9]{inputenc}
\usepackage{amsmath}
\usepackage{amssymb}
\usepackage{graphicx}
\usepackage{wasysym}

\makeatletter

\usepackage{hyperref}
\usepackage{color}

\makeatother

\usepackage{babel}
\begin{document}
\title{Operator entanglement in $\mathrm{SU}(2)$-symmetric dissipative quantum many-body dynamics}
\author{Lin Zhang}
\thanks{lin.zhang@icfo.eu}
\affiliation{ICFO-Institut de Ciencies Fotoniques, The Barcelona Institute of Science and Technology, Av. Carl Friedrich Gauss 3, 08860 Castelldefels (Barcelona), Spain}
\begin{abstract}
The presence of symmetries can lead to nontrivial dynamics of operator entanglement in open quantum many-body systems, which characterizes the cost of an matrix product density operator (MPDO) representation of the density matrix in the tensor-network methods and provides insights into the corresponding classical simulability. One example is the $\mathrm{U}(1)$-symmetric open quantum systems with dephasing, in which the operator entanglement increases logarithmically at late times instead of being suppressed by the dephasing. Here we numerically study the far-from-equilibrium dynamics of operator entanglement in a dissipative quantum many-body system with the more complicated $\mathrm{SU}(2)$ symmetry and dissipations beyond dephasing. We show that after the initial rise and fall, the operator entanglement also increases again in a logarithmic manner at late times in the $\mathrm{SU}(2)$-symmetric case. We find that this behavior can be fully understood from the corresponding $\mathrm{U}(1)$ subsymmetry by considering the symmetry-resolved operator entanglement. Especially, the probability distribution of different $\mathrm{U}(1)$ sectors also follows the Gaussian distribution observed in the $\mathrm{U}(1)$-symmetric case with dephasing, with the variance growing in a power law. But unlike the latter, both the classical Shannon entropy associated with the probabilities for the half system being in different symmetry sectors and the corresponding symmetry-resolved operator entanglement now have nontrivial contributions to the late time logarithmic growth of operator entanglement in our $\mathrm{SU}(2)$-symmetric case. Our results show evidence that the logarithmic growth of operator entanglement at long times is a generic behavior of dissipative quantum many-body dynamics with $\mathrm{U}(1)$ as the symmetry or subsymmetry and for more broad dissipations beyond dephasing, although more analytical or numerical proofs are still required for this conjecture in the future studies. By breaking the $\mathrm{SU}(2)$ symmetry of our quantum many-body dynamics to $\mathrm{U}(1)$, we also show that the latter property---the logarithmic growth behavior of operator entanglement for more broad dissipations beyond dephasing---is valid even for open quantum systems with only $\mathrm{U}(1)$ symmetry.
\end{abstract}
\maketitle
\section{Introduction}

Recent years have witnessed remarkable experimental advances in atomic, molecular, and optical physics, which have enabled us to engineer quantum many-body systems in controllable and clean environments at the level of individual atoms, molecules, and ions~\cite{Bloch2012,Blatt2012,Gadway2016,Browaeys2020,Monroe2021}. These achievements provide us insights into the studies of strongly correlated quantum systems~\cite{Lewenstein2007,Bloch2008,Dutta2015,Aidelsburger2021}. Among them, extensive attention has been focused on understanding quantum many-body dynamics far away from equilibrium~\cite{Polkovnikov2011,Eckardt2017}, an outstanding challenge in the modern physical sciences, and many new states of matter have been uncovered, such as the many-body localization~\cite{Schreiber2015,Smith2016}, time crystals~\cite{Choi2017,Zhang2017}, and quantum many-body scars~\cite{Bernien2017,Turner2018}. However, it is usually hard to characterize these nonequilibrium systems. 

One important tool to describe the quantum many-body dynamics is the so-called entanglement entropy. On the one hand, its growth as a function of time can show fundamentally distinct behavior in different nonequilibrium systems. For instance, the growth of entanglement entropy is logarithmic in the many-body localized models~\cite{Abanin2019}, while it increases linearly for a generic chaotic system. This provides a defining feature for many nonequilibrium states of matter.  On the other hand, the growth of entanglement entropy also provides insights into the classical simulations of quantum many-body dynamics using tensor networks~\cite{Verstraete2008,Schollwoeck2011,Paeckel2019}. In one dimension (1D), the quantum states can be faithfully represented by matrix-product states (MPSs) in terms of local rank-three tensors with bond dimension $\chi$~\cite{Vidal2004,Perez-Garcia2007}, for which the entanglement entropy is bounded by $\log_{2}\chi$. Thus the linear increase of entanglement entropy will lead to an exponential growth of bond dimension, which is considered to be hard to simulate~\cite{Schuch2008}. Only the quantum dynamics within a low-entanglement manifold of the Hilbert space can be efficiently simulated on classical computers. Thanks to the above reasons, studying the growth of entanglement entropy has become one of the central tasks in investigating quantum many-body dynamics and has attracted broad interest~\cite{Calabrese2005,Fagotti2008,Alba2017,Lukin2019}.

The growth of entanglement entropy also helps the understanding of open quantum many-body systems, which are ubiquitous in practical experiments due to the inevitable couplings to environments. Like the MPS representation for pure states, the density matrix of 1D open quantum many-body systems can be described by the matrix product density operators (MPDOs)~\cite{Zwolak2004,Verstraete2004,Werner2016}. The bipartition of MPDO through Schmidt decomposition further defines the entanglement entropy in operator space, which is known as the operator entanglement and characterizes the cost of an MPDO representation, i.e., how many Schmidt values are needed at least for faithfully representing an operator~\cite{Zanardi2000,Zanardi2001,Wang2002,Prosen2007,Pizorn2009,Dubail2017}, hence providing insights into the classical simulability of open quantum many-body systems via the MPDO tensor-network method.

In open quantum systems, after the initial linear growth reminiscent of unitary quantum dynamics, the operator entanglement is expected to be suppressed by the dissipations. However, it was recently reported that the operator entanglement in $\mathrm{U}(1)$-symmetric open quantum many-body systems with dephasing increases logarithmically at long times~\cite{Wellnitz2022}, which is attributed to the growth of classical Shannon entropy associated with the probabilities for the half system being in different $\mathrm{U}(1)$ sectors. This highlights the role of symmetries on the operator entanglement dynamics in open quantum many-body systems, which, however, still remains largely unexplored. One important symmetry is the non-Abelian $\mathrm{SU}(2)$ symmetry, which is believed to be crucial in many nonequilibrium phenomena like quantum many-body scars~\cite{Choi2019} and anomalous finite-temperature transport~\cite{Ilievski2021}. As the non-Abelian symmetry in general increases the entanglement entropy~\cite{Majidy2023,Li2024,Moharramipour2024}, it would be interesting to study how the long-time behavior of operator entanglement is impacted in the presence of $\mathrm{SU}(2)$ symmetry. Particularly, as the $\mathrm{SU}(2)$-symmetric open quantum many-body system contains $\mathrm{U}(1)$ as the subsymmetry and dissipations beyond dephasing, it would be interesting to know whether the logarithmic growth of operator entanglement can go beyond open quantum systems with $\mathrm{U}(1)$ symmetry and dephasing.

In this work, we numerically study the far-from-equilibrium dynamics of operator entanglement in a dissipative quantum many-body system with $\mathrm{SU}(2)$ symmetry. We show that after the initial rise and fall, the operator entanglement also increases again in a logarithmic manner at long times in the $\mathrm{SU}(2)$-symmetric case. We find that this behavior can be fully understood from the corresponding $\mathrm{U}(1)$ subsymmetry by considering the symmetry-resolved operator entanglement~\cite{Li2023,Rath2023,Murciano2023}. Especially, the probability distribution of different $\mathrm{U}(1)$ sectors also follows the Gaussian distribution observed in the $\mathrm{U}(1)$-symmetric case with dephasing~\cite{Wellnitz2022}, with the variance growing in a power law. But unlike the latter, both the classical Shannon entropy and the symmetry-resolved operator entanglement now have nontrivial contributions to the late-time logarithmic growth of operator entanglement in our $\mathrm{SU}(2)$-symmetric dissipative quantum many-body dynamics. Our results show evidence that the logarithmic growth of operator entanglement at long times is a generic behavior of dissipative quantum many-body dynamics with $\mathrm{U}(1)$ as the symmetry or subsymmetry and for more broad dissipations beyond dephasing, although more analytical or numerical proofs are still required for this conjecture in the future studies. By breaking the $\mathrm{SU}(2)$ symmetry of our quantum many-body dynamics to $\mathrm{U}(1)$, we also show that the latter property---the logarithmic growth behavior of operator entanglement for more broad dissipations beyond dephasing---is valid even for open quantum systems with only $\mathrm{U}(1)$ symmetry.

The remaining part of this article is organized as follows. In Sec.~\ref{sec:model}, we introduce the dissipative quantum many-body model studied in this work. Then in Sec.~\ref{sec:MPDO decomposition of density matrix}, we describe the MPDO decomposition of density matrix. After introducing the (symmetry-resolved) operator entanglement in Sec.~\ref{sec:operator entanglement}, we present the numerical results in Sec.~\ref{sec:results}. Finally, the conclusion is provided in Sec.~\ref{sec:conclusion}.

\section{Model\label{sec:model}}

\begin{figure}
\includegraphics{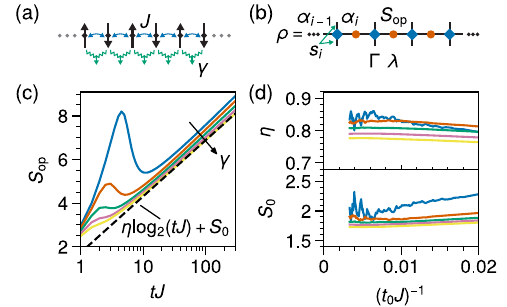}
\caption{Operator entanglement dynamics in the $\mathrm{SU}(2)$-symmetric dissipative quantum many-body system. (a) Sketch of the model. We consider a quantum spin chain with coherent nearest-neighbor coupling $J$ (blue arrows) and local dissipation proportional to the dipole interaction between neighbor sites with strength $\gamma$ (green arrows). (b) MPDO decomposition of the density matrix $\rho$ in terms of local tensors $\Gamma$ (blue squares) and $\lambda$ (orange circles). Here $\alpha_{i-1}$ and $\alpha_{i}$ are the bond indices, while $s_{i}$ denotes the combined physical index of the bra and ket legs at site $i$. The operator entanglement $S_{\mathrm{op}}$ at certain bond can be calculated from the Schmidt vectors $\lambda$. (c) Time evolution of $S_{\mathrm{op}}$ for the product initial state of singlet pairs with different dissipation strength $\gamma=0.05J$, $0.10J$, $0.15J$, $0.20J$, and $0.25J$. The black dashed line indicates the logarithmic growth of operator entanglement at long times (log-scale time axis), i.e., $S_{\mathrm{op}}(t\to\infty)=\eta\log_{2}(tJ)+S_{0}$. (d) Numerical prefactor $\eta$ and offset $S_{0}$ obtained from the local tangent of operator entanglement at time $t_{0}J$. The results are converged for time step $\delta tJ=0.5$ and maximum bond dimension $\chi=50000$. \protect\label{fig:operator entanglement dynamics in SU(2)-symmetric open many-body quantum system}}
\end{figure}

We consider the open quantum many-body dynamics on an infinite spin-$1/2$ chain governed by the Lindblad master equation ($\hbar\equiv1$)
\begin{equation}
\frac{\mathrm{d}}{\mathrm{d}t}\rho=-\mathrm{i}[H,\rho]+\gamma\sum_{i}\left(L_{i}\rho L_{i}^{\dagger}-\{L_{i}^{\dagger}L_{i},\rho\}/2\right)\equiv\mathcal{L}[\rho],\label{eq:Lindblad master equation}
\end{equation}
where the Hamiltonian $H=J\sum_{i}P_{i,i+1}$ with the exchange operator $P_{i,i+1}=2\mathbf{S}_{i}\cdot\mathbf{S}_{i+1}+1/2$ is the $1$D Heisenberg model with $\mathrm{SU}(2)$ symmetry. Here $\mathbf{S}_{i}$ denotes the spin-$1/2$ operator on site $i$, and $J$ is the nearest-neighbor spin coupling strength. We couple the system to an environment through the Lindblad operator $L_{i}=P_{i,i+1}$ with strength $\gamma$, which describes the dissipation proportional to the dipole interaction between neighbor sites; see Fig.~\ref{fig:operator entanglement dynamics in SU(2)-symmetric open many-body quantum system}(a) for the sketch. Similar model has been considered in Refs.~\cite{DeNardis2021,Claeys2022} to study the stability or absence of superdiffusion in a spin chain with fluctuating exchange couplings that break the integrability. Since $L_{i}$ commutes with the total spin operator, the above Lindblad master equation preserves the $\mathrm{SU}(2)$ symmetry. To study the quantum dynamics with $\mathrm{SU}(2)$ symmetry, we consider the initial state $\rho_{0}=\vert\psi_{0}\rangle\langle\psi_{0}\vert$ with $\vert\psi_{0}\rangle=\bigotimes_{i}(\left|\uparrow\right\rangle _{2i-1}\left|\downarrow\right\rangle _{2i}-\left|\downarrow\right\rangle _{2i-1}\left|\uparrow\right\rangle _{2i})/\sqrt{2}$, which is the product state of singlet pairs. This state has total spin 0 and is $\mathrm{SU}(2)$-symmetric. 

\section{MPDO decomposition of density matrix\label{sec:MPDO decomposition of density matrix}}

To solve the Lindblad master equation, we describe the density matrix with MPDO. The matrix product decomposition of density matrix is a mixed-state version of the MPS, where the density matrix $\rho$ of a spin-$1/2$ chain with local Hilbert space dimension $d=2$ is treated as a vector $\vert\rho\rangle_{\sharp}$ in the tensor product space of the $d\times d$ complex matrices~\cite{Zwolak2004}. Here we use the subscript $\sharp$ (sharp) to denote operators (superoperators) when represented as the superkets $\vert\cdot\rangle_{\sharp}$ (mappings between superkets), with $\langle s\vert s'\rangle_{\sharp}\equiv\mathrm{Tr}(s^{\dagger}s')/d$. Given a set of orthonormal basis $\{\vert s_{i}\rangle_{\sharp}\}$ for each site $i$, the density matrix $\vert\rho\rangle_{\sharp}$ can be decomposed as
\begin{equation}
\vert\rho\rangle_{\sharp}=\sum_{\{s_{i}\}}\sum_{\{\alpha_{i}\}}\prod_{i}\Gamma_{\alpha_{i-1}\alpha_{i}}^{[i]s_{i}}\lambda_{\alpha_{i}}^{[i]}\bigotimes_{i}\vert s_{i}\rangle_{\sharp}\label{eq:MPDO decomposition}
\end{equation}
through a succession of Schmidt decompositions~\cite{Zwolak2004}, where $\Gamma^{[i]}$ is a rank-three tensor on site $i$ with one combined physical index $s_{i}$ and two bond indices $\alpha_{i-1}$, $\alpha_{i}$, while $\lambda^{[i]}$ is the Schmidt vector on bond $i$; see Fig.~\ref{fig:operator entanglement dynamics in SU(2)-symmetric open many-body quantum system}(b). Since both the initial state and Liouvillian operator are invariant under the translation by two sites, we have $\Gamma^{[i+2]}=\Gamma^{[i]}$ and $\lambda^{[i+2]}=\lambda^{[i]}$. Therefore, the numerical simulation can be performed within a unit cell of two sites for the infinite lattice, for which only two $\Gamma$ and $\lambda$ tensors are needed to capture the density matrix.

For the Liouvillian superoperator \eqref{eq:Lindblad master equation} that can be decomposed into terms involving at most two contiguous sites, the time evolution can be simulated using the infinite time-evolving block decimation (iTEBD) algorithm with a fourth-order Trotter decomposition of the matrix exponential of the Liouvillian $\exp(\mathcal{L}_{\sharp}\delta t)$ for a time step $\delta t$~\cite{Paeckel2019}. We note that since the open quantum many-body dynamics governed by the Lindblad master equation is nonunitary, the canonical form of the infinite MPDO is in general not preserved in the original iTEBD algorithm~\cite{Vidal2007}. As a consequence, the truncation errors accumulate fast and ruin the simulation at long times. To overcome this issue, we employ the improved algorithm for nonunitary evolutions proposed in Ref.~\cite{Orus2008}, in which the tensors updated in each time step are reorthonormalized into the canonical form. To avoid the exponential growth of bond dimension, we also truncate the tensors $\{\Gamma^{[i]}\}$ and $\{\lambda^{[i]}\}$ at a maximum bond dimension $\chi$. The results shown in this work are all numerically converged; see Appendix~\ref{sec:numerical convergence}.

\section{Operator entanglement\label{sec:operator entanglement}}

We focus on the time evolution of operator entanglement in our $\mathrm{SU}(2)$-symmetric dissipative quantum many-body dynamics. The operator entanglement is a basis-independent measure for quantum operators. Given the MPDO decomposition \eqref{eq:MPDO decomposition}, the operator entanglement at certain bond is defined as~\cite{Zanardi2000,Zanardi2001,Wang2002,Prosen2007,Pizorn2009,Dubail2017}
\begin{equation}
S_{\mathrm{op}}\equiv-\sum_{\alpha}\lambda_{\alpha}^{2}\log_{2}\lambda_{\alpha}^{2},
\end{equation}
where the Schmidt values are assumed to be normalized, i.e., $\sum_{\alpha}\lambda_{\alpha}^{2}=1$. We note that the operator entanglement does not characterize the quantum entanglement between different parts of the system, which is instead captured by the entanglement measures like entanglement negativities when $\rho$ is a mixed state~\cite{Vidal2002,Plenio2005}. The operator entanglement mainly reflects the cost of encoding an operator in the MPDO representation and also provides insights into nonequilibrium quantum many-body physics like the quantum chaos and information scrambling~\cite{Zhou2017,Alba2019,Styliaris2021}.

For our $\mathrm{SU}(2)$-symmetric quantum dynamics, the total spin $\mathbf{S}_{\mathrm{tot}}=\sum_{i}\mathbf{S}_{i}$ is conserved. Therefore, we have $\mathbf{S}_{\mathrm{tot}}^{2}\rho=S_{\mathrm{tot}}(S_{\mathrm{tot}}+1)\rho$ during the time evolution, where $S_{\mathrm{tot}}$ is zero for our initial state $\rho_{0}$ as a product state of singlet pairs. Consider a bipartitioning of the system, $\rho=\sum_{\alpha}\lambda_{\alpha}\varrho_{\alpha}^{[\mathrm{L}]}\otimes\varrho_{\alpha}^{[\mathrm{R}]}$, with $\langle\varrho_{\alpha}^{[\mathrm{L/R}]}\vert\varrho_{\alpha'}^{[\mathrm{L/R}]}\rangle_{\sharp}=\delta_{\alpha\alpha'}$ for the left (L) and right (R) part. We also have $\mathbf{S}_{\mathrm{L/R}}^{2}\varrho_{\alpha}^{[\mathrm{L/R}]}=S_{\mathrm{L/R}}(S_{\mathrm{L/R}}+1)\varrho_{\alpha}^{[\mathrm{L/R}]}$ with $S_{\mathrm{L}}=S_{\mathrm{R}}\equiv S$, where $\mathbf{S}_{\mathrm{L/R}}$ is the total spin operator in the left/right part of the system. For this, we can relabel the bond index $\alpha$ as $\alpha\to(S,i_{S})$, where $i_{S}$ distinguishes the Schmidt values (i.e., different half-system density matrices) corresponding to the same half-system total spin $S$. We note that in our notation the degenerate degrees of freedom of each spin sector $S$ is not counted into $i_{S}$. For each bond label $(S,i_{S})$, there are actually $(2S+1)$ Schmidt coefficients with the same value $\lambda_{(S,i_{S})}$ in our numerical simulation with $\mathrm{SU}(2)$ symmetry. Hence the operator entanglement is given by $S_{\mathrm{op}}=-\sum_{S}\sum_{i_{S}}(2S+1)\lambda_{(S,i_{S})}^{2}\log_{2}\lambda_{(S,i_{S})}^{2}$ in this notation with the normalization condition $\sum_{S}\sum_{i_{S}}(2S+1)\lambda_{(S,i_{S})}^{2}=1$. We also remark that the quantum number $S$ can also be defined by multiplying the symmetry operator from the right of density matrix, which does not change the results.

It is useful to also introduce the symmetry-resolved operator entanglement, which has attracted extensive attention recently~\cite{Wellnitz2022,Li2023,Rath2023,Murciano2023}. In this work, we follow the definition proposed in Refs.~\cite{Wellnitz2022,Li2023}, while alternative definitions are also possible~\cite{Rath2023,Murciano2023}. Specifically, for a generic global symmetry $Q$, the operator entanglement for the MPDO decomposition \eqref{eq:MPDO decomposition} at certain bond can be split into different symmetry sectors, similar to that of the state entanglement~\cite{Laflorencie2014,Goldstein2018,Xavier2018,Barghathi2018,Barghathi2019,Parez2021}. The symmetry-resolved operator entanglement in a specific symmetry sector $q$ is defined through the renormalized Schmidt coefficients $\hat{\lambda}_{q,i_{q}}=\lambda_{q,i_{q}}/\sqrt{p_{q}}$ belonging to that sector via $S_{\mathrm{op},q}\equiv-\sum_{i_{q}}\hat{\lambda}^2_{q,i_{q}}\log_{2}\hat{\lambda}^2_{q,i_{q}}$, where $i_{q}$ labels different Schmidt coefficients in the symmetry sector $q$ and $p_{q}=\sum_{i_{q}}\lambda^2_{q,i_{q}}$ is the probability of having symmetry charge $q$ in the half system such that $\sum_{i_{q}}\hat{\lambda}^2_{q,i_{q}}=1$. In our $\mathrm{SU}(2)$-symmetric case, different symmetry sectors are labeled by the half-system total spin $S$, and the probability of having total spin $S$ in the half system is $p_{S}=(2S+1)\sum_{i_{S}}\lambda_{(S,i_{S})}^{2}$ [note that in our notation for the $\mathrm{SU}(2)$ symmetry, each bond label $(S,i_{S})$ actually corespondents to $(2S+1)$ Schmidt coefficients with the same value $\lambda_{S,i_{S}}$]. Then the symmetry-resolved operator entanglement in the spin sector $S$ is given by $S_{\mathrm{op},S}=-(2S+1)\sum_{i_{S}}\hat{\lambda}_{(S,i_{S})}^{2}\log_{2}\hat{\lambda}_{(S,i_{S})}^{2}$ with $\hat{\lambda}_{(S,i_{S})}\equiv\lambda_{(S,i_{S})}/\sqrt{p_{S}}$. With this, the full operator entanglement can be recast as
\begin{equation}
S_{\mathrm{op}}=\sum_{S}p_{S}S_{\mathrm{op},S}-\sum_{S}p_{S}\log_{2}p_{S},\label{eq:relation between the full and symmetry-resolved operator entanglement}
\end{equation}
where the first term describes the averaged contribution from the symmetry-resolved operator entanglement in different spin sectors, while the second term is the classical Shannon entropy of the probability distribution. 

Since our model has the $\mathrm{U}(1)$ subsymmetry, e.g., the conservation of total magnetization along the $z$ direction, we can also label the Schmidt coefficients via the half-system magnetization and define the corresponding symmetry-resolved operator entanglement; see Refs.~\cite{Wellnitz2022,Rath2023,Murciano2023}. For our $\mathrm{SU}(2)$-symmetric case, since each spin sector with $S\geq|S_{z}|$ contributes one state to the magnetization sector $S_{z}$, we can express the probability of having magnetization $S_{z}$ in the half system as $p_{S_{z}}=\sum_{S\geq|S_{z}|}\sum_{i_{S}}\lambda_{(S,i_{S})}^{2}$, and the symmetry-resolved operator entanglement in this magnetization sector is given by $S_{\mathrm{op},S_{z}}=-\sum_{S\geq|S_{z}|}\sum_{i_{S}}\tilde{\lambda}_{(S,i_{S})}^{2}\log_{2}\tilde{\lambda}_{(S,i_{S})}^{2}$ with $\tilde{\lambda}_{(S,i_{S})}\equiv\lambda_{(S,i_{S})}/\sqrt{p_{S_{z}}}$. The relation to the full operator entanglement $S_{\mathrm{op}}$ is similar to Eq.~\eqref{eq:relation between the full and symmetry-resolved operator entanglement}.

\begin{figure}
\includegraphics{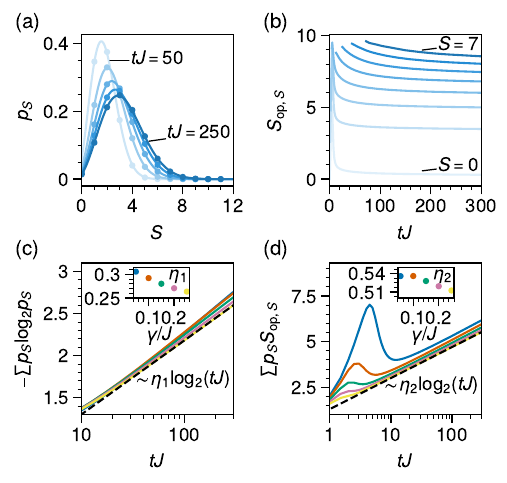}
\caption{Symmetry-resolved operator entanglement in spin sectors. (a) Probabilities $p_{S}$ of the infinite chain with total spin $S$ in the half system at increasingly late times ($50\protect\leq tJ\protect\leq250$ from light to dark) for $\gamma=0.05J$. The lines are fits with the trial function shown in the main text. (b) The corresponding symmetry-resolved operator entanglement $S_{\mathrm{op},S}$ as a function of time. Only the data with probabilities $p_{S}>10^{-4}$ are presented. (c) (d) The classical Shannon entropy $-\sum p_{S}\log_{2}p_{S}$ and the averaged symmetry-resolved operator entanglement $\sum p_{S}S_{\mathrm{op},S}$ as a function of time for $\gamma/J=0.05$, $0.10$, $0.15$, $0.20$, and $0.25$. The black dashed lines indicate the logarithmic growth at late times (log-scale time axis). The inserts present the corresponding prefactors. The results are converged for the time step $\delta tJ=0.5$ and maximum bond dimension $\chi=50000$.\protect\label{fig:symmetry-resolved operator entanglement in spin sectors}}
\end{figure}

\section{Results\label{sec:results}}

\subsection{Logarithmic growth of operator entanglement}

We now present the numerical results for our open quantum many-body system~\eqref{eq:Lindblad master equation}. In Fig.~\ref{fig:operator entanglement dynamics in SU(2)-symmetric open many-body quantum system}(c), we show the time evolution of operator entanglement $S_{\mathrm{op}}$ for the product initial state of singlet pairs. Initially,  the operator entanglement grows
linearly in time, as the quantum dynamics at such short times is dominated by the Hamiltonian part of Eq.~\eqref{eq:Lindblad master equation} and can be approximated to be unitary. However, when $t\apprge\gamma^{-1}/4$ in our numerical simulation, the coupling to the environment becomes relevant, and the operator entanglement starts to decrease.

Typically, if there is no conservation law in the dissipative quantum many-body dynamics, the density matrix is expected to relax to the infinite-temperature state (i.e., the identity matrix), and the operator entanglement converges toward the corresponding value of that stationary state at late times. However, the introduction of symmetries can enrich the behavior of operator entanglement. Particularly, it was recently reported that after the initial rise and fall, the operator entanglement in $\mathrm{U}(1)$-symmetric open quantum many-body systems with dephasing increases again in a logarithmic manner at late times, i.e., $S_{\mathrm{op}}(t\to\infty)=\eta\log_{2}(tJ)+S_{0}$~\cite{Wellnitz2022}. 

Here we also observe the same behavior of operator entanglement in our dissipative quantum many-body dynamics with $\mathrm{SU}(2)$ symmetry; see Fig.~\ref{fig:operator entanglement dynamics in SU(2)-symmetric open many-body quantum system}(c). Especially, we show the prefactor $\eta$ and offset $S_{0}$ as a function of time $t_{0}$ in Fig.~\ref{fig:operator entanglement dynamics in SU(2)-symmetric open many-body quantum system}(d), which is obtained as the local tangent of operator entanglement. Due to the limited maximum bond dimension, the data for small $\gamma$ is not good enough as those for the strongly dissipative cases. However, all of them show the tendency to converge to a finite value as $t_{0}\to\infty$, thus identifying the logarithmic growth behavior of operator entanglement at late times. We note that unlike the $\mathrm{U}(1)$-symmetric case with dephasing~\cite{Wellnitz2022}, from the numerical tendency exhibited in Fig.~\ref{fig:operator entanglement dynamics in SU(2)-symmetric open many-body quantum system}(d) both the prefactor $\eta$ and offset $S_{0}$ seem to be nonuniversal and depend on the value of the dissipation strength. However, it should be mentioned that whether $\eta$ converges universally to a fixed value like Ref.~\cite{Wellnitz2022} or not in the limit $t\to\infty$ is still an open question due to the limited time achieved in our numerical simulation.

\subsection{Symmetry-resolved operator entanglement in spin sectors}

To understand the logarithmic growth behavior of operator entanglement in the quantum dynamics with $\mathrm{SU}(2)$ symmetry, it is useful to consider the symmetry-resolved operator entanglement. We first consider the spin sectors. We show the probabilities $p_{S}$ of having total spin $S$ in the half system at late times in Fig.~\ref{fig:symmetry-resolved operator entanglement in spin sectors}(a). The half-system total spin is mainly distributed in the low-spin sectors, but as time increases the system has more probabilities in the high-spin sectors. We fit the probability distribution with the trial function 
\begin{equation}\label{eq:probability distribution among different spin sectors}
    p_{S}=\frac{2S+1}{\sqrt{2\pi\delta^{2}}}\left[e^{-S^{2}/2\delta^{2}}-e^{-(S+1)^{2}/2\delta^{2}}\right]
\end{equation}
with parameter $\delta$ and find a good match {[}see lines in Fig.~\ref{fig:symmetry-resolved operator entanglement in spin sectors}(a){]}. Considering the relation $p_{S}=(2S+1)(p_{S_{z}=S}-p_{S_{z}=S+1})$, this suggests that the probability of having magnetization $S_{z}$ in the half system follows the Gaussian distribution with variance $\delta$ in our dissipative quantum many-body dynamics with $\mathrm{SU}(2)$ symmetry, like the one observed in the $\mathrm{U}(1)$-symmetric case with dephasing~\cite{Wellnitz2022}. 

We also present the symmetry-resolved operator entanglement in different spin sectors in Fig.~\ref{fig:symmetry-resolved operator entanglement in spin sectors}(b). Unlike Ref.~\cite{Wellnitz2022}, where the symmetry-resolved operator entanglement drops to very small values quickly, here the operator entanglement $S_{\mathrm{op},S}$ remains finite and large at late times, indicating that in addition to the classical Shannon entropy of probability distributions, the symmetry-resolved operator entanglement also have nontrivial contributions to the logarithmic growth of the total operator entanglement in our $\mathrm{SU}(2)$-symmetric case, as we show in Figs.~\ref{fig:symmetry-resolved operator entanglement in spin sectors}(c) and \ref{fig:symmetry-resolved operator entanglement in spin sectors}(d). The nonvanishing contribution of symmetry-resolved operator entanglement at late times is compatible with the results shown in Refs.~\cite{Li2024,Moharramipour2024}, in which the authors studied the bipartite entanglement of the stationary states for unital quantum channels on a finite chain at $t\to\infty$ and found that the half-system operator entanglement scales logarithmically with the system size for the open quantum many-body systems with $\mathrm{SU}(2)$ symmetry. In this $t\to\infty$ limit, it was shown that the total operator entanglement involves both quantum and classical contributions; see Eq.~(23) in Ref.~\cite{Moharramipour2024} or Eq.~(20) in Ref.~\cite{Li2024}. The former is related to the nontrivial Hilbert space structure with dimension $2S+1>1$ for each spin sectors $S>0$, while the latter is attributed to the probability distribution of states among different spin sectors. Both the quantum and classical contributions were shown to scale logarithmically with the system size and diverge in the thermodynamic limit, which is compatible with our results that both the classical Shannon entropy of probability distributions and the averaged symmetry-resolved operator entanglement for an infinite spin chain grow logarithmically with the time and also diverge as $t\to\infty$. Therefore, it would be expected that the nonvanishing contribution of symmetry-resolved operator entanglement at late times in our $\mathrm{SU}(2)$-symmetric case arises from the nontrivial spin sector structures. We note that although the operator entanglement shows a logarithmic behavior both in system size in the $t\to\infty$ limit and in time in the limit of infinite system size, it seems that these two phenomena do not have too much connection with each other as the prefactor $\eta$ in our work is different from those shown in Refs.~\cite{Li2024,Moharramipour2024}.

\begin{figure}
\includegraphics{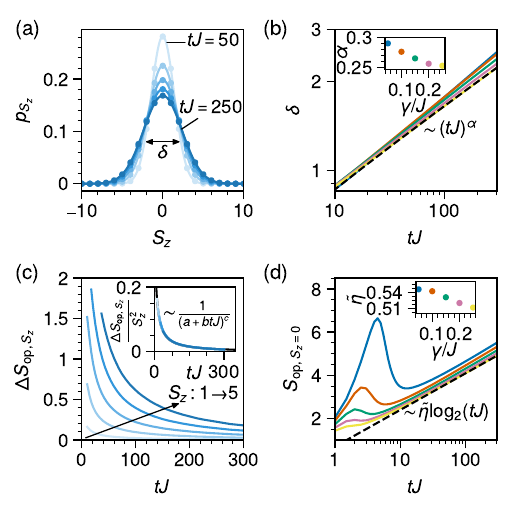}
\caption{Symmetry-resolved operator entanglement in magnetization sectors. (a) Probabilities $p_{S_{z}}$ of the infinite chain with magnetization $S_{z}$ in the half system at increasingly late times ($50\protect\leq tJ\protect\leq250$ from light to dark) for $\gamma=0.05J$. Lines are the Gaussian fits. (b) Variance $\delta$ of the Gaussian fits as a function of time for $\gamma/J=0.05$, $0.10$, $0.15$, $0.20$, and $0.25$. The black dashed line indicates $\delta\sim(tJ)^{\alpha}$ at late times (double-log scale). The corresponding exponent $\alpha$ for each $\gamma$ is shown in the insert. (c) Symmetry-resolved operator entanglement difference $\Delta S_{\mathrm{op},S_{z}}$ as a function of time for $\gamma=0.05J$ and $S_{z}=1,2,3,4,5$ (light to dark). The insert shows the data scaled by $1/S_{z}^{2}$, and the black line is the fit $(a+btJ)^{-c}$ with $a=2.4964$, $b=0.2554$, and $c=1.1228$. (d) Symmetry-resolved operator entanglement $S_{\mathrm{op},S_{z}}$ in the magnetization sector $S_{z}=0$ as a function of time for various $\gamma$. The black dashed line indicates the logarithmic growth of $S_{\mathrm{op},S_{z}=0}$ at late times (log-scale time axis), with the corresponding prefactors shown in the insert. Here the results are converged for time step $\delta tJ=0.5$ and maximum bond dimension $\chi=50000$.\protect\label{fig:symmetry-resolved operator entanglement in magnetization sectors}}
\end{figure}

\subsection{Symmetry-resolved operator entanglement in magnetization sectors}

Despite the different behavior of symmetry-resolved operator entanglement at late times, the analysis of probabilities $p_{S}$ suggests that the logarithmic growth behavior of operator entanglement in our dissipative quantum many-body dynamics with $\mathrm{SU}(2)$ symmetry can also be understood from the corresponding $\mathrm{U}(1)$ subsymmetry by considering the symmetry-resolved operator entanglement in magnetization sectors. 

As we expected, the probabilities $p_{S_{z}}$ indeed are approximately Gaussian at late times, i.e., $p_{S_{z}}\simeq e^{-S_{z}^{2}/2\delta^{2}}/\sqrt{2\pi\delta^{2}}$; see Fig.~\ref{fig:symmetry-resolved operator entanglement in magnetization sectors}(a). In Fig.~\ref{fig:symmetry-resolved operator entanglement in magnetization sectors}(b), we also present the time dependence of the variance $\delta$, which follows the power law $\sim(tJ)^{\alpha}$ at late times and explains the logarithmic growth behavior of $-\sum_{S}p_{S}\log_{2}p_{S}$ shown in  Fig.~\ref{fig:symmetry-resolved operator entanglement in spin sectors}(c). The exponent $\alpha$ in general depends on the dissipation strength and for large $\gamma$ is close to the value $0.25$ predicted for the $\mathrm{U}(1)$-symmetric dissipative quantum dynamics with dephasing~\cite{Wellnitz2022}; see the insert in Fig.~\ref{fig:symmetry-resolved operator entanglement in magnetization sectors}(b). Obviously, the classical Shannon entropy of the probabilities in magnetization sectors, $-\sum_{S_{z}}p_{S_{z}}\log_{2}p_{S_{z}}\simeq\log_{2}\delta+\log_{2}\sqrt{2\pi e}$, which follows a logarithmical growth law with prefactor $\alpha$, also cannot capture the whole prefactor $\eta$ of the logarithmic growth of operator entanglement at late times; cf. Fig.~\ref{fig:operator entanglement dynamics in SU(2)-symmetric open many-body quantum system}(d).

To identify the behavior of $S_{\mathrm{op},S_{z}}$, we first consider the difference of symmetry-resolved operator entanglement between the finite magnetization sector and $S_{z}=0$ sector by defining $\Delta S_{\mathrm{op},S_{z}}\equiv S_{\mathrm{op},S_{z}}-S_{\mathrm{op},S_{z}=0}$. The results for $\gamma=0.05J$ are presented in Fig.~\ref{fig:symmetry-resolved operator entanglement in magnetization sectors}(c), where as the time increases $\Delta S_{\mathrm{op},S_{z}}$ decays to very small values. This suggests that the symmetry-resolved operator entanglement $S_{\mathrm{op},S_{z}}$ will take the same value at long times for all of the magnetization sectors. In Appendix~\ref{sec:proof of the late-time same value behavior}, a proof for this observation is provided, based on the properties of symmetry-resolved operator entanglement $S_{\mathrm{op},S}$ and probability distribution $p_{S}$ in the spin sectors. Although these properties are obtained from numerical simulations, our proof indeed indicates that the late-time same value behavior of $S_{\mathrm{op},S_{z}}$ arises from the $\mathrm{SU}(2)$ symmetry of the dissipative quantum many-body dynamics. Moreover, we find that $\Delta S_{\mathrm{op},S_{z}}$ scaled by $1/S_{z}^{2}$ collapses almost onto each other at late times for various magnetization sector $S_{z}$ and can be captured by the function $(a+btJ)^{-c}$ with $c\approx1$; see the insert in Fig.~\ref{fig:symmetry-resolved operator entanglement in magnetization sectors}(c). As the contributions of the symmetry-resolved operator entanglement can be rewritten as $\sum_{S_{z}}p_{S_{z}}S_{\mathrm{op},S_{z}}=S_{\mathrm{op},S_{z}=0}+\sum_{S_{z}\neq0}p_{S_{z}}\Delta S_{\mathrm{op},S_{z}}$, where the second term $\sim(tJ)^{2\alpha-c}$ vanishes at long times, this suggests that the late-time behavior of the averaged symmetry-resolved operator entanglement can be captured by $S_{\mathrm{op},S_{z}=0}$, which, just like the total operator entanglement $S_{\mathrm{op}}$, after the initial rise and fall, increases again in a logarithmic manner at late times; see Fig.~\ref{fig:symmetry-resolved operator entanglement in magnetization sectors}(d). Together with the classical Shannon entropy of the probability distribution $p_{S_{z}}$, this explains the long-time behavior of the total operator entanglement in our $\mathrm{SU}(2)$-symmetric dissipative quantum dynamics. We note that in our $\mathrm{SU}(2)$-symmetric case, the symmetry-resolved operator entanglement contributes a logarithmic term with larger prefactor $\eta$ than the classical Shannon entropy of probability distributions (see Fig.~\ref{fig:symmetry-resolved operator entanglement in spin sectors} and Fig.~\ref{fig:symmetry-resolved operator entanglement in magnetization sectors}), while in the $\mathrm{U}(1)$-symmetric case with dephasing considered in Ref.~\cite{Wellnitz2022} the logarithmic growth of Shannon entropy has a larger prefactor (indeed the only nontrivial term).

Since the operator entanglement dynamics in our $\mathrm{SU}(2)$-symmetric case can be fully understood from the corresponding $\mathrm{U}(1)$ subsymmetry, the above results show evidence that the logarithmic growth of operator entanglement at long times is a generic behavior of the dissipative quantum many-body dynamics with $\mathrm{U}(1)$ as the symmetry or subsymmetry, although more analytical or numerical proofs are still required for this conjecture in the future studies. The logarithmic growth behavior of operator entanglement also holds for more broad dissipations beyond dephasing. In the following, by breaking the symmetry of our quantum dynamics to $\mathrm{U}(1)$, we show that this property is valid even for the open quantum systems with only $\mathrm{U}(1)$ symmetry.

\subsection{Symmetry breaking from SU(2) to U(1)}\label{subsec:Symmetry breaking from SU(2) to U(1)}

\begin{figure}
\includegraphics{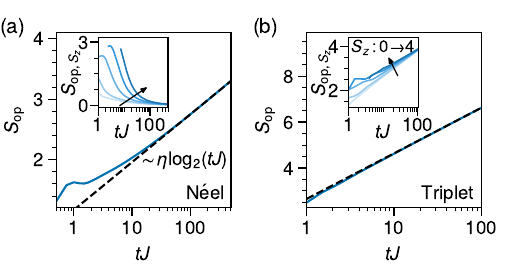}
\caption{Operator entanglement for quantum dynamics with symmetry being broken to $\mathrm{U}(1)$. We consider the N{\' e}el initial state in (a) and the product initial state of triplet pairs in (b). The black dashed lines indicate the logarithmic growth of operator entanglement at late times (log-scale time axis), with the prefactor $\eta\approx 0.24$ for (a) and $\approx0.60$ for (b), respectively. The inserts show the time evolution of symmetry-resolved operator entanglement $S_{\mathrm{op},S_{z}}$ for $S_{z}=0,1,2,3,4$ (from light to dark). Here we set $\gamma=0.5J$. The time step $\delta tJ$ is $0.25$ for (a) and $0.5$ for (b). We choose the maximum bond dimension $\chi$ as $4000$ for (a) and $12000$ for (b). \protect\label{fig:operator entanglement for quantum dynamics with symmetry being broken to U(1)}}
\end{figure}

We consider the quantum dynamics starting from the N{\' e}el state $\vert\psi_{0}\rangle=\bigotimes_{i}\left|\uparrow\right\rangle _{2i-1}\left|\downarrow\right\rangle _{2i}$ and the product state of triplet pairs $\vert\psi_{0}\rangle=\bigotimes_{i}(\left|\uparrow\right\rangle _{2i-1}\left|\downarrow\right\rangle _{2i}+\left|\downarrow\right\rangle _{2i-1}\left|\uparrow\right\rangle _{2i})/\sqrt{2}$, which break the $\mathrm{SU}(2)$ symmetry to $\mathrm{U}(1)$ at the level of initial states. The results for dissipation strength $\gamma=0.5J$ are presented in Fig.~\ref{fig:operator entanglement for quantum dynamics with symmetry being broken to U(1)}. For both initial states, the operator entanglement $S_{\mathrm{op}}$ shows the logarithmic growth behavior at late times, although the symmetry-resolved operator entanglement $S_{\mathrm{op},S_{z}}$ decays to small values at late times for the N{\' e}el initial state [see the insert of Fig.~\ref{fig:operator entanglement for quantum dynamics with symmetry being broken to U(1)}(a)] while it increases logarithmically for the product initial state of triplet pairs as in the $\mathrm{SU}(2)$-symmetric case [see the insert of Fig.~\ref{fig:operator entanglement for quantum dynamics with symmetry being broken to U(1)}(b)]. Since the dissipation in our model \eqref{eq:Lindblad master equation} is proportional to the dipole interaction between neighbor sites, this demonstrates that the logarithmic growth behavior of operator entanglement at late times holds for more broad dissipations beyond dephasing even for the open quantum many-body dynamics with only $\mathrm{U}(1)$ symmetry. 

We note that it would be interesting to further explore the conditions in which the late-time symmetry-resolved operator entanglement exhibits nontrivial growth or vanishes when the symmetry of dissipative quantum many-body dynamics is broken to $\mathrm{U}(1)$. However, as suggested by Fig.~\ref{fig:operator entanglement for quantum dynamics with symmetry being broken to U(1)}, this question highly depends on the initial states and probably also on the terms introduced to break the $\mathrm{SU}(2)$ symmetry of Liouvillian to $\mathrm{U}(1)$. Since we cannot explore all these possibilities in the numerical simulations, we leave this question for future studies.

\section{Conclusion\label{sec:conclusion}}

In conclusion, we have studied the far-from-equilibrium dynamics of operator entanglement in a dissipative quantum many-body system with $\mathrm{SU}(2)$ symmetry. We find that after the initial rise and fall, the operator entanglement increases again in a logarithmic manner at late times. This behavior can be fully understood from the corresponding $\mathrm{U}(1)$ subsymmetry by considering the symmetry-resolved operator entanglement. Especially, the probability distribution of different $\mathrm{U}(1)$ sectors also follows the Gaussian distribution observed in the $\mathrm{U}(1)$-symmetric case with dephasing, with the variance growing in a power law. But unlike the latter, in addition to the classical Shannon entropy associated with the probabilities for the half system being in different symmetry sectors, the symmetry-resolved operator entanglement also contributes nontrivially to the late-time growth of operator entanglement in our $\mathrm{SU}(2)$-symmetric case. Our results show numerical evidence that the logarithmic growth of operator entanglement at late times is a generic behavior of dissipative quantum many-body dynamics with $\mathrm{U}(1)$ as the symmetry or subsymmetry and for more broad dissipations beyond dephasing. In the future studies, it would be interesting to further test this conjecture by investigating more symmetries like $\mathrm{SU}(N>2)$, $\mathrm{SO}(N)$, and $\mathrm{SP}(2N)$, and provide more analytical proofs. Moreover, in addition to the strong symmetries considered so far, it would be meaningful to study how the presence of weak symmetries impacts the entanglement dynamics of open quantum systems~\cite{Buca2012,Kawabata2023}. 

\begin{acknowledgments}
We implement the $\mathrm{SU}(2)$ symmetry of the tensors using the package \texttt{TensorKit.jl}~\citep{TensorKit}. We acknowledge support from: European Research Council AdG NOQIA; MCIN/AEI (PGC2018-0910.13039/501100011033, CEX2019-000910-S/10.13039/501100011033, Plan National FIDEUA PID2019-106901GB-I00, Plan National STAMEENA PID2022-139099NB, I00, project funded by MCIN/AEI/10.13039/501100011033 and by the ``European Union NextGenerationEU/PRTR'' (PRTR-C17.I1), FPI); QUANTERA DYNAMITE PCI2022-132919, QuantERA II Programme co-funded by European Union's Horizon 2020 program under Grant Agreement No 101017733; Ministry for Digital Transformation and of Civil Service of the Spanish Government through the QUANTUM ENIA project call - Quantum Spain project, and by the European Union through the Recovery, Transformation and Resilience Plan - NextGenerationEU within the framework of the Digital Spain 2026 Agenda; MICIU/AEI/10.13039/501100011033 and EU (PCI2025-163167); Fundaci{\' o} Cellex; Fundaci{\' o} Mir-Puig; Generalitat de Catalunya (European Social Fund FEDER and CERCA program); Barcelona Supercomputing Center MareNostrum (FI-2023-3-0024); Funded by the European Union. Views and opinions expressed are however those of the author(s) only and do not necessarily reflect those of the European Union, European Commission, European Climate, Infrastructure and Environment Executive Agency (CINEA), or any other granting authority. Neither the European Union nor any granting authority can be held responsible for them (HORIZON-CL4-2022-QUANTUM-02-SGA PASQuanS2.1, 101113690, EU Horizon 2020 FET-OPEN OPTOlogic, Grant No 899794, QU-ATTO, 101168628), EU Horizon Europe Program (This project has received funding from the European Union's Horizon Europe research and innovation program under Grant Agreement No. 101080086 --- NeQST); ICFO Internal ``QuantumGaudi'' project.
\end{acknowledgments}

\section*{Data availability}

The data that support the findings of this article are not publicly available. The data are available from the authors upon reasonable request.

\appendix

\section{Details on numerical convergence\label{sec:numerical convergence}}

\begin{figure}
\includegraphics{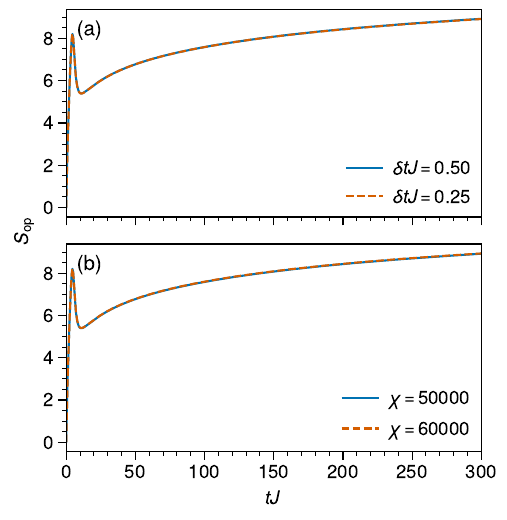}
\caption{Numerical convergence of operator entanglement $S_{\mathrm{op}}$. (a) Convergence in the time step $\delta t$. (b) Convergence in the maximum bond dimension $\chi$. Here we set $\gamma=0.05J$. The maximum bond dimension $\chi$ in (a) is $50000$, and the time step $\delta tJ$ in (b) is $0.5$.\protect\label{fig:numerical convergence}}
\end{figure}

In this Appendix, we provide more details on the numerical convergence of operator entanglement $S_{\mathrm{op}}$ in time step $\delta t$ and maximum bond dimension $\chi$.

We use the iTEBD algorithm with a Trotter decomposition of the matrix exponential of superoperator $\mathcal{L}_{\sharp}$ for each time step $\delta t$ to solve the dissipative quantum many-body dynamics. In order to reach long times, the fourth-order Trotter decomposition~\cite{Paeckel2019} is employed:
\begin{equation}
    e^{\mathcal{L}_{\sharp}\delta t}=U(\delta t_{1})U(\delta t_{2})U(\delta t_{3})U(\delta t_{2})U(\delta t_{1})
\end{equation}
with
\begin{equation}
    U(\delta t_{i}) = e^{\mathcal{L}_{\sharp,\mathrm{odd}}\delta t_{i}/2}e^{\mathcal{L}_{\sharp,\mathrm{even}}\delta t_{i}}e^{\mathcal{L}_{\sharp,\mathrm{odd}}\delta t_{i}/2}
\end{equation}
and
\begin{equation}
    \delta t_{1} = \delta t_{2} = \frac{1}{4-4^{1/3}}\delta t,\qquad\delta t_{3}=\delta t - 2\delta t_{1} - 2\delta t_{2}.
\end{equation}
Here $\mathcal{L}_{\sharp,\mathrm{even/odd}}$ is the superoperator acting on the even/odd bonds. 

As we show in Fig.~\ref{fig:numerical convergence}(a), this method allows for the numerical results converged for time steps even up to $\delta tJ=0.5$. In Fig.~\ref{fig:numerical convergence}(b), we also present the operator entanglement $S_{\mathrm{op}}$ for $\gamma=0.05J$ with maximum bond dimension $\chi=50000$ and $60000$. The visually indistinguishable lines indicate that the results are already well converged at $\chi=50000$ up to time $tJ=300$. For larger $\gamma=0.10J$, $0.15J$, $0.20J$, and $0.25J$, the operator entanglement will be reduced. Therefore, the maximal bond dimension $\chi=50000$ is enough to guarantee the numerical convergence for these $\gamma$'s.

We note that in the plots of prefactor $\eta$ and offset $S_{0}$ of $S_{\mathrm{op}}(t\to\infty)=\eta\log_{2}(tJ)+S_{0}$ shown in Figs.~\ref{fig:operator entanglement dynamics in SU(2)-symmetric open many-body quantum system}(c) and \ref{fig:operator entanglement dynamics in SU(2)-symmetric open many-body quantum system}(d), which are obtained by fitting the data set $\{(\log_{2}(tJ),S_{\mathrm{op}}(t))\}$ for $t = t_{0},t_{0}\pm\delta t$ via a linear function, the results for small $\gamma=0.05J$ and $\gamma=0.10J$ exhibit an artificial wiggle behavior at late times. This arises from numerical errors due to the finite time step and maximal bond dimension. Especially, the latter has a more significant influence as the artificial wiggle behavior is highly reduced when the dissipation strength $\gamma$ increases, which reduces the operator entanglement and increases the accuracy of our numerical simulation with fixed maximal bond dimension $\chi=50000$. To remove this artificial wiggle behavior of $\eta$ and $S_{0}$ for these small $\gamma$'s, we could choose a smaller time step and/or a larger maximal bond dimension to further increase the accuracy of our numerical simulations, which, however, requires much more expensive computational resources. Moreover, as we show in the main text, most properties of the operator entanglement, e.g., the late-time logarithmic growth and probability distribution of different symmetry sectors, have already been captured very well in our current numerical simulation. This artificial wiggle behavior at small $\gamma$'s does not affect our results.

\section{Proof of the late-time same value behavior of $S_{\mathrm{op},S_{z}}$\label{sec:proof of the late-time same value behavior}}

In this Appendix, we provide a proof for the observation that the symmetry-resolved operator entanglement $S_{\mathrm{op},S_{z}}$ takes the same value at long times for all of the magnetization sectors, based on the properties of symmetry-resolved operator entanglement $S_{\mathrm{op},S}$ and probability distribution $p_{S}$ in the spin sectors that (i) the total operator entanglement $S_{\mathrm{op}}$ grows logarithmically at late times, i.e., $S_{\mathrm{op}}\sim\log_{2}(tJ)$; (ii) the probability distribution $p_{S}$ among different spin sectors follows Eq.~\eqref{eq:probability distribution among different spin sectors} with $\delta\sim(tJ)^{\alpha}$ at late times [see Fig.~\ref{fig:symmetry-resolved operator entanglement in spin sectors}(a) and Fig.~\ref{fig:symmetry-resolved operator entanglement in magnetization sectors}(b)]; (iii) $S_{\mathrm{op},S}$ remains finite at late times [see Fig.~\ref{fig:symmetry-resolved operator entanglement in spin sectors}(b)].

\begin{widetext}
We focus on the region with $S_{z}\geq 0$. The proof for the cases with $S_{z}<0$ is similar. According to the expressions for $S_{\mathrm{op},S}$ ($S_{\mathrm{op},S_{z}}$) and $p_{S}$ ($p_{S_{z}}$) in terms of Schmidt values $\lambda_{S,i_{S}}$ (see Sec.~\ref{sec:operator entanglement}), we find the following relations
\begin{align}
p_{S_{z}} & =\sum_{S\geq S_{z}}\frac{p_{S}}{2S+1}=\frac{1}{\sqrt{2\pi\delta^{2}}}e^{-S_{z}^{2}/2\delta^{2}},\\
S_{\mathrm{op},S_{z}} & =\frac{1}{p_{S_{z}}}\sum_{S\geq S_{z}}\frac{p_{S}(S_{\mathrm{op},S}-\log_{2}p_{S})}{2S+1}+\log_{2}p_{S_{z}.}
\end{align}
We consider the difference bewteen $S_{\mathrm{op},S_{z}}$ and $S_{\mathrm{op},S_{z}-1}$:
\begin{equation}
\begin{aligned} & S_{\mathrm{op},S_{z}}-S_{\mathrm{op},S_{z}-1}\\
= & \left(\frac{1}{p_{S_{z}}}-\frac{1}{p_{S_{z}-1}}\right)\sum_{S\geq S_{z}}\frac{p_{S}(S_{\mathrm{op},S}-\log_{2}p_{S})}{2S+1}-\frac{1}{p_{S_{z}-1}}\frac{p_{S=S_{z}-1}(S_{\mathrm{op},S=S_{z}-1}-\log_{2}p_{S=S_{z}-1})}{2S_{z}-1}+\log_{2}\frac{p_{S_{z}}}{p_{S_{z}-1}}\\
= & \left(\frac{1}{p_{S_{z}}}-\frac{1}{p_{S_{z}-1}}\right)\sum_{S\geq S_{z}}\frac{p_{S}(S_{\mathrm{op},S}-\log_{2}p_{S})}{2S+1}-\left(1-\frac{p_{S_{z}}}{p_{S_{1}-1}}\right)(S_{\mathrm{op},S=S_{z}-1}-\log_{2}p_{S=S_{z}-1})+\log_{2}\frac{p_{S_{z}}}{p_{S_{z}-1}}\\
\leq & \left(\frac{1}{p_{S_{z}}}-\frac{1}{p_{S_{z}-1}}\right)\underbrace{\sum_{S}p_{S}(S_{\mathrm{op},S}-\log_{2}p_{S})}_{S_{\mathrm{op}}}-\left(1-\frac{p_{S_{z}}}{p_{S_{1}-1}}\right)(S_{\mathrm{op},S=S_{z}-1}-\log_{2}p_{S=S_{z}-1})+\log_{2}\frac{p_{S_{z}}}{p_{S_{z}-1}}.
\end{aligned}
\end{equation}
For the first term, we have 
\begin{equation}
    \left(\frac{1}{p_{S_{Z}}}-\frac{1}{p_{S_{z}-1}}\right)S_{\mathrm{op}} = \sqrt{2\pi\delta^{2}}\left[e^{S_{z}^{2}/2\delta^{2}}-e^{(S_{z}-1)^{2}/2\delta^{2}}\right]S_{\mathrm{op}}\simeq \sqrt{2\pi\delta^{2}}e^{S_{z}^{2}/2\delta^{2}}(2S_{z}-1)S_{\mathrm{op}}/2\delta^{2}\sim \frac{\log_{2}(tJ)}{(tJ)^{\alpha}},
\end{equation}
which tends to zero at late times. For the second term, on the one hand, $(1-p_{S_{Z}}/p_{S_{z}-1})S_{\mathrm{op},S=S_{z}-1}\simeq(2S_{z}-1)S_{\mathrm{op},S=S_{z}-1}/2\delta^{2}$ vanishes as $t\to\infty$ since $S_{\mathrm{op},S}$ remains finite at late times. On the other hand, we have
\begin{equation}
\begin{aligned} & \left(1-p_{S_{z}}/p_{S_{z}-1}\right)\log_{2}p_{S=S_{z}-1}\\
= & \left(1-p_{S_{z}}/p_{S_{z}-1}\right)\log_{2}\left(\frac{2S_{z}-1}{\sqrt{2\pi\delta^{2}}}e^{-S_{z}^{2}/2\delta^{2}}[e^{(2S_{z}-1)/2\delta^{2}}-1]\right)\simeq\frac{2S_{z}-1}{2\delta^{2}}\log_{2}\left[\frac{(2S_{z}-1)^{2}}{\sqrt{8\pi\delta^{6}}}\right]\sim\frac{\log_{2}\delta}{\delta^{2}}\sim\frac{\log_{2}(tJ)}{(tJ)^{2\alpha}},
\end{aligned}
\end{equation}
which also vanishes at late times. Finally, the third term $\log_{2}(p_{S_{z}}/p_{S_{z}-1})=\log_{2}e^{-(2S_{z}-1)/2\delta^{2}}$ tends to zero at late times. Therefore, we have proved that $S_{\mathrm{op},S_{z}}-S_{\mathrm{op},S_{z}-1}$ vanishes as $t\to\infty$ and $S_{\mathrm{op},S_{z}}$ takes the same value at late times for all $S_{z}$.

We note that although the properties of $S_{\mathrm{op},S}$ and $p_{S}$ are obtained from numerical simulations, the above proofs indeed indicate that the late-time same value behavior of $S_{\mathrm{op},S_{z}}$ arises from the $\mathrm{SU}(2)$ symmetry of the dissipative quantum many-body dynamics.
\end{widetext}

% \bibliography{operator_entanglement_xxx_su2}


\end{document}